\documentclass[twocolumn,aps,pra,superscriptaddress,a4paper,sort&compress,balancelastpage]{revtex4-2}
\usepackage[normalem]{ulem}
\usepackage{amsmath}
\usepackage{amssymb}
\usepackage{gensymb}
\usepackage{bm}
\usepackage{graphicx}
\usepackage{float}
\usepackage{subfigure}
\usepackage[dvipsnames]{xcolor}
\usepackage{placeins}
\usepackage{braket}
\usepackage{bbold}
\usepackage{ulem}
\usepackage[colorlinks, linkcolor=blue, citecolor=blue, urlcolor=blue, breaklinks=red]{hyperref}

\usepackage{bbm}

\usepackage{lipsum}
\usepackage{physics}
\usepackage{dsfont}

\definecolor{Mycolor1}{HTML}{44aa99}
\definecolor{Mycolor2}{HTML}{cc6677}
\usepackage{orcidlink}

\begin{document}

\title{Physically Accessible and Inaccessible Quantum Correlations of Dirac Fields in Schwarzschild Spacetime}

\author{Samira Elghaayda~\!\!\orcidlink{0000-0002-6655-0465}}
\email{samira.elghaayda-etu@etu.univh2c.ma}
\affiliation{ Laboratory of High Energy Physics and Condensed Matter, Department of Physics,\\ Faculty of Sciences of Aïn Chock, Hassan II University,\\ P.O. Box 5366 Maarif, Casablanca 20100, Morocco.}

\author{Asad Ali\orcidlink{0000-0001-9243-417X}} \email{asal68826@hbku.edu.qa}
\affiliation{Qatar Centre for Quantum Computing, College of Science and Engineering, Hamad Bin Khalifa University, Qatar Foundation, Doha, Qatar.}

\author{Saif Al-Kuwari\orcidlink{0000-0002-4402-7710}}
\email{smallkuwari@hbku.edu.qa}
\affiliation{Qatar Centre for Quantum Computing, College of Science and Engineering, Hamad Bin Khalifa University, Qatar Foundation, Doha, Qatar.}

\author{Mostafa Mansour~\!\!\orcidlink{0000-0003-0821-0582}}
\email{mostafa.mansour.fsac@gmail.com}
\affiliation{ Laboratory of High Energy Physics and Condensed Matter, Department of Physics,\\ Faculty of Sciences of Aïn Chock, Hassan II University,\\ P.O. Box 5366 Maarif, Casablanca 20100, Morocco.}

\begin{abstract}
In this study, we investigate the influence of Hawking decoherence on the quantum correlations of Dirac fields between \textit{Alice} and \textit{Bob}. Initially, they share a \textit{Gisin} state near the Schwarzschild black hole (SBH) in an asymptotically flat region. Then, \textit{Alice} remains stationary in this region, while \textit{Bob} hovers near the event horizon (EH) of the SBH. We expect that \textit{Bob}, using his excited detector, will detect a thermal Fermi-Dirac particle distribution. We assess the quantum correlations in the evolved \textit{Gisin} state using quantum consonance and uncertainty-induced non-locality across physically accessible, physically inaccessible, and spacetime regions. Our investigation examines how these measures vary with Hawking temperature, Dirac particle frequency, and the parameters of the initial \textit{Gisin} state. Additionally, we analyze the distribution of these quantum correlation measures across all possible regions, noting a redistribution towards the physically inaccessible region. Our findings demonstrate that Hawking decoherence reduces the quantum correlations of Dirac fields in the physically accessible region, with the extent of reduction depending on the initial state parameters. Moreover, as Hawking decoherence intensifies in the physically inaccessible and spacetime regions, the quantum correlations of Dirac fields reemerge and ultimately converge to specific values at infinite Hawking temperature. These results contribute to our understanding of quantum correlation dynamics within the framework of relativistic quantum information (RQI).
\end{abstract}


\maketitle


\section{Introduction}

From Einstein's perspective, the gravitational collapse of sufficiently massive stars gives rise to intriguing black holes (BHs), which stand as captivating entities in our universe \cite{schwarzschild1916gravitationsfeld,hawking1976breakdown}. According to the no-hair theorem, BHs preserve only information concerning their mass, charge, and angular momentum, while concealing all other specifics \cite{gurlebeck2015no}. Recent advancements in astronomy have provided both direct and indirect evidence of these enigmatic objects. Notably, the detection of gravitational waves by advanced LIGO and Virgo detectors marked a significant milestone, revealing the merger of binary BH systems \cite{abbott2016observation}. Additionally, the Event Horizon Telescope (EHT) achieved a monumental feat by capturing the first-ever image of a supermassive BH at the center of the massive elliptical galaxy $M87$ \cite{collaboration2019first,akiyama2019first2,akiyama2019first3,akiyama2019first4,akiyama2019first5,akiyama2019first6}. Furthermore, EHT also photographed $Sgr$ $A^*$, another supermassive BH \cite{akiyama2022first7}. Simulations have shown growing interest in modeling Hawking radiation from BHs and generating cosmological particles in quantum systems \cite{tian2017analog,drori2019observation,viermann2022quantum,steinhauer2022analogue}. The Hawking effect has generated significant interest, resulting in extensive research that investigates its implications in various physical contexts \cite{peres2004,friis2012,bruschi2013,martin2013,martin2014,wu2022genuine,bhuvaneswari2022,chowdhury202}.

Modern physics is built upon the foundational theories of quantum mechanics and general relativity. However, the challenge of unifying these theories remains. The emerging field of RQI seeks to bridge this gap \cite{peres2004}. Through its exploration, RQI not only enhances our understanding of quantum mechanics but also provides new insights into long-standing puzzles, such as the \textit{BH information paradox}. This paradox revolves around the unitary evaporation of BHs \cite{hawking2015information,hawking1976breakdown,hawking1975particle,hawking19}. Understanding how relativistic particles become entangled in both flat and curved spacetimes is not only intriguing for quantum information but also holds significant implications for BH physics \cite{shi2004entanglement, friis2010relativistic, friis2012, bruschi2013, martin2013, martin2014}. Extracting information about entanglement from quantum field vacua has thus become crucial for comprehending both the vacuum itself and the underlying spacetime structure \cite{benatti2004t, zhou2021entanglement}. This interest has spurred research into quantum information within BH backgrounds and the dynamics of entanglement \cite{wu2022genuine,bhuvaneswari2022, chowdhury202, aspachs2010, unruh1976, crispino2008, dewitt1967, brown2013, chung2019black, ali2024quantum, elghaayda2023entropy, elghaayda2024distribution}. For instance, \textit{Pan et al.} explored entanglement in scalar fields near BHs \cite{pan2008}, while \textit{Deng et al.} investigated the effect of Hawking decoherence and prepared states on the distillability of entanglement in Dirac fields \cite{deng2011hawking}. \textit{Xu et al.} extended this study to analyze the impact of Hawking radiation on multipartite entanglement in Schwarzschild spacetime (SST) \cite{xu2014hawking}. Furthermore, Hawking effect-induced entanglement degradation has been shown to increase the entropic uncertainty bound \cite{feng2015uncertainty}. Recently, \textit{Pan et al.} achieved a significant milestone in the study of RQI by using the "Micius" quantum communication satellite to experimentally test gravitational decoherence effects \cite{xu2019satellite}.

Quantum resource theories provide a foundational framework for understanding the essential characteristics of physical systems, encompassing non-locality \cite{popescu19, brunner20, elghaayda20}, quantum coherence \cite{streltsov201, baumgratz201, oumennana2022quantum}, entanglement \cite{vedral2002role, horodecki200, wootters20, man01, man02}, and quantum correlations \cite{ollivier200, girolami201, mansour2021quantum}. By offering an operational interpretation of these quantum phenomena, this theory illuminates the mechanisms underlying complex systems \cite{sbiri, elghaayda2023quantum, elghaayda202}. The Hawking effect in BHs has been observed to consistently degrade the quantum resources of quantum fields in curved spacetime \cite{wu2023genuine}, akin to environmental decoherence that constrains certain quantum information processes. Therefore, gaining a quantitative understanding of this decoherence in non-inertial frames is crucial for discussing specific quantum information tasks. While the distribution of quantum correlations in a relativistic framework has been explored within the realm of RQI, there are still unanswered questions that require further investigation. We aim to explore whether the Hawking decoherence uniformly diminishes the quantum correlations of Dirac fields, and if its impact varies across different measures of quantum correlations. To investigate this, we analyze the dynamics of quantum correlations of Dirac fields in distinct regions of spacetime — accessible, inaccessible, and spacetime boundaries. Our evaluation employs the uncertainty-induced non-locality measure \cite{wu2014uncertainty, elghaayda20, elghaayda2024enhancing} to assess quantum correlations and utilizes the concept of quantum consonance to quantify both separable and entangled correlations \cite{pei2012using, motavallibashi2021, khedif2021thermal, elghaayda2023entropy}.

Through meticulous analysis, we investigate how Hawking decoherence affects the dynamics of quantum correlations among Dirac fields across different regions: accessible, inaccessible, and spacetime. Our study reveals that Hawking decoherence consistently diminishes these correlations in the accessible region. Surprisingly, in inaccessible regions, these correlations experience a resurgence as Hawking decoherence intensifies. This resurgence persists until the correlations asymptotically approach certain constants at extremely high Hawking temperatures. This phenomenon stems from the formation of quantum correlations across the event horizon for Dirac fields. Finally, we explore the distribution of these correlations among subsystems, aiming to comprehensively grasp the evolving dynamics of quantum correlations among Dirac fields.

The rest of this paper is organized as follows: Section \ref{sec2} provides the definitions of the metrics used to assess quantum correlations of Dirac fields within accessible, inaccessible, and spacetime regions. Next, in Section \ref{model}, we revisit the vacuum structure and Hawking decoherence for Dirac fields in SST. By using the Kruskal basis formulation, we derive a pure state that is shared between particles inside and outside the EH. The findings and subsequent discussions are presented in Section \ref{sec4}, and concluding remarks are provided in Section \ref{sec5}.

\section{Quantum correlations measures} \label{sec2}
In this section, we provide a brief overview of the fundamental concepts of quantum correlations in the current system. We will specifically focus on quantum consonance and uncertainty-induced nonlocality, as these are commonly used as quantifiers for measuring quantum correlations.

\subsection{Quantum consonance}
Quantum consonance is a quantum measurement that integrates entanglement and other quantum correlations by eliminating local coherence from global coherence using local unitary operations \cite{pei2012using, motavallibashi2021, khedif2021thermal}. It serves as a metric for quantifying quantum correlations, mathematically represented as a composite of entanglement and specific other quantum correlations \cite{pei2012using}. This formulation arises from applying local unitary transformations that isolate the local coherence contribution from the overall coherence.
\begin{align}
	\mathcal{C_{AB}}(\varrho)=\sum_{ijmn}\left| \varrho_{ijmn}^{q}(1-\delta_{im})(1-\delta_{jn})\right|,
	\label{eq1}
\end{align}
The state $\varrho^{q} = (\mathcal{W}_1 \otimes \mathcal{W}_2)\varrho(\mathcal{W}_1 \otimes \mathcal{W}_2)^{\dagger}$ is derived by applying local unitary operations $\mathcal{W}_1$ and $\mathcal{W}_2$ to the original state $\varrho$. For a two-qubit X-state, it's noteworthy that quantum consonance equals the sum of the off-diagonal elements of the density operator.

\subsection{Uncertainty-induced nonlocality}
The concept of uncertainty-induced nonlocality, as introduced in \cite{wu2014uncertainty}, quantifies nonlocal quantum correlations differently from measurement-induced nonlocality (MIN). MIN is originally defined as the maximal change in a bipartite quantum state due to a projective measurement, but suffers from non-contractivity issues and is absent in states lacking MIN. Initially, MIN is computed as the maximum Hilbert-Schmidt distance between the quantum state and its post-measurement counterpart on a subsystem, affecting the state globally without altering local reduced density operators \cite{luo2011measurement, hu2015measurement}. In contrast, uncertainty-induced nonlocality is defined as the maximum skew information, and serves as a revised measure that resolves these issues and assesses the maximum skew information correlations of any bipartite state $\varrho$ \cite{wu2014uncertainty}.
\begin{equation}
	\mathcal{U_{AB}}(\varrho)=\max_{\zeta^{C}}\mathcal{I}(\varrho, \zeta^{C}),
	\label{eq2}
\end{equation}
where,
\begin{equation}
	\mathcal{I}(\varrho, \zeta)=-\frac{1}{2}\mathrm{Tr}\left[ \sqrt{\varrho}, \zeta\right]^{2},
	\label{eq3}
\end{equation}
Skew information, as introduced by Wigner and Yanase \cite{wigner1963information, luo2003wigner}, quantifies the uncertainty resulting from the observable $\zeta$ acting on the subsystem $\varrho$. In Eq. (\ref{eq2}), the maximization is performed over all locally maximally informative commuting observables $\zeta^{C}=\zeta_{A}^{C}\otimes \mathbb{I}_{B}$, where $\zeta^{C}_{A}$ is a Hermitian operator acting on qubit $A$ with distinct eigenvalues, and $\mathbb{I}_{B}$ denotes the identity operator on qubit $B$. The precise form of the uncertainty-induced nonlocality is explicitly detailed for any ($2\otimes d$)-dimensional quantum state $\varrho$ in \cite{wu2014uncertainty}.

\begin{equation}
	\mathcal{U_{AB}}(\varrho)=\left\lbrace \begin{array}{cc}
		1-n_{min}(\mathcal{N}), & \quad \pmb{v}=\pmb{0}  \\
		1-\frac{1}{|\pmb{v}|^{2}}\,\pmb{v}\, \mathcal{N}\,\pmb{v}^{T}, & \quad  \pmb{v} \neq \pmb{0}
	\end{array}\right.
	\label{eq4}
\end{equation}
The notation $\pmb{v}^{T}$ denotes the transpose of the Bloch vector $\pmb{v}$, and $n_{\min}(\mathcal{N})$ represents the smallest eigenvalue of the symmetric real matrix $\mathcal{N}_{3\times 3}$ defined by its elements.
\begin{equation}
	(\mathcal{N})_{ij}=\mathrm{Tr}\left\lbrace \varrho^{1/2}(\hat{\sigma}_{i}^{A}\otimes \mathbb{I}_{B})\varrho ^{1/2}(\hat{\sigma}_{j}^{A}\otimes \mathbb{I}_{B})\right\rbrace,
	\label{eq5}
\end{equation}
where $\{\hat{\sigma}_{i}^{A}\}_{i = x, y, z}$ represents the conventional Pauli matrices that operate on qubit $A$.
\section{\textit{Gisin} state in SST \label{model}}
In this part, we aim to delve into the concepts of fermionic fields and vacuum structure in SST, particularly from the perspective of observers of an initial \textit{Gisin} state. This state notably manifests complete entanglement for an observer in free fall towards a SBH.
\subsection{\textit{Gisin} state}
\textit{Gisin} states, introduced by Gisin in \cite{Gisin1996hidden}, represent a class of states notable for their unique properties. These states encompass combinations of pure, entangled, and separable mixed states, and have been instrumental in demonstrating hidden nonlocality. Specifically, certain \textit{Gisin} states initially exhibit local behavior but can lose this characteristic after undergoing purely local operations \cite{Gisin1996hidden, gs1, gs2, sbiri}. Mathematically, the \textit{Gisin} states can be expressed as:
\begin{equation}
	\varrho_{\lambda,\psi}(0)= \lambda | \phi_{\psi}\rangle \langle \phi_{\psi}| + \frac{1 - \lambda}{2} \left( |00\rangle\langle00| + |11\rangle\langle 11| \right),
	\label{eq19}
\end{equation}
The mixing parameter $\lambda \in [0, 1]$ defines the degree of mixture, while the state $|\phi_{\psi}\rangle = \sin(\psi)|01\rangle + \cos(\psi)|10\rangle$ represents a Bell-like state with $\psi \in [0, \pi/2]$. At $\psi = \frac{\pi}{4}$, $|\phi_{\psi}\rangle$ becomes the familiar Bell state $|\phi^{+}\rangle = \frac{1}{\sqrt{2}}(|01\rangle + |10\rangle)$. Gisin states are distinct from Weyl states but intersect non-trivially when locally maximally mixed at $\psi = \frac{\pi}{4}$. For $\lambda < 1$, the Gisin state is fully mixed. However, at $\lambda = 1$, it simplifies to the pure entangled state $\left| \phi_{\psi}\right\rangle \left\langle \phi_{\psi}\right|$. To understand how $\lambda$ and $\psi$ influence the correlation measures $\mathcal{U_{AB}}$ and $\mathcal{C_{AB}}$, we present the results in Figure (\ref{figure01}) below.

\begin{figure}[H]
	\centering
	\subfigure[]{\label{fig01}\includegraphics[scale=0.6]{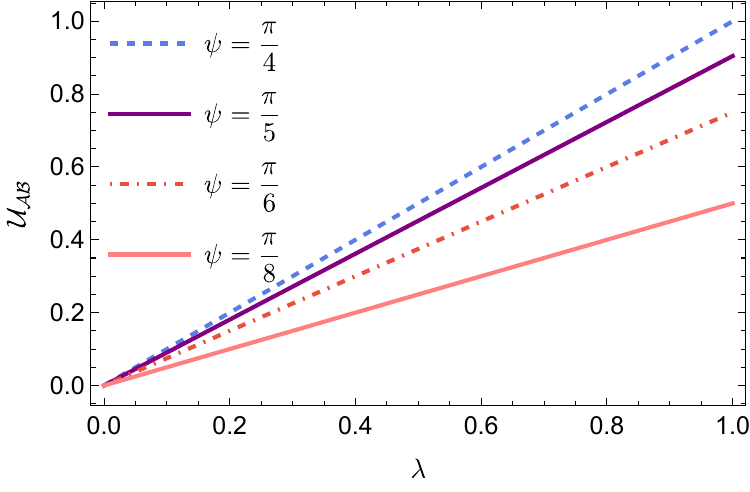}}
	\subfigure[]{\label{fig02}\includegraphics[scale=0.6]{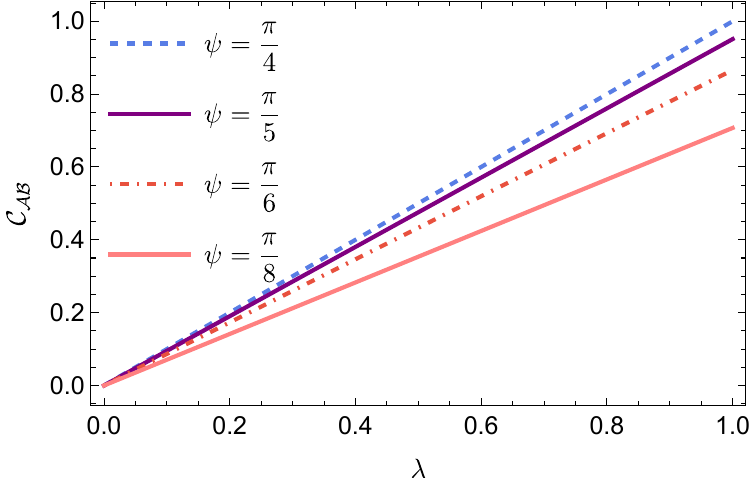}}
	\caption{Quantum correlations $\mathcal{U_{AB}}$ \ref{fig01} and $\mathcal{C_{AB}}$ \ref{fig02} as a function of the mixing parameter $\lambda$ for different angle values $\psi$ for the initial \textit{Gisin} state $\varrho_{\lambda,\psi}(0)$. }
	\label{figure01}
\end{figure}

From Figure (\ref{figure01}), it is evident that both $\mathcal{U_{AB}}$ and $\mathcal{C_{AB}}$ consistently increase with the mixing parameter. Initially, $\mathcal{C_{AB}}$ and $\mathcal{U_{AB}}$ share a common minimum value of $\mathcal{U_{AB}} = \mathcal{C_{AB}} = 0$, due to the separable and incoherent nature of the \textit{Gisin} state when $\lambda \approx 0$. Both measures then steadily grow, reaching their maximum degree of correlation at $\lambda = 1$, signifying a transition from a partially correlated state to a maximally correlated state. Additionally, it is notable that $\mathcal{U_{AB}}(\psi = \frac{\pi}{4}) > \mathcal{U_{AB}}(\psi = \frac{\pi}{5})$ and $\mathcal{C_{AB}}(\psi = \frac{\pi}{4}) > \mathcal{C_{AB}}(\psi = \frac{\pi}{5})$. This implies that the \textit{Gisin} state exhibits the highest level of quantum correlations when $\psi = \frac{\pi}{4}$. Interestingly, for $\psi < \frac{\pi}{4}$ and $\lambda = 1$, $\mathcal{C_{AB}}$ bounds the $\mathcal{U_{AB}}$ quantifier. This indicates that the parameters $\psi$ and $\lambda$ significantly influence the behavior of the $\mathcal{U_{AB}}$ and $\mathcal{C_{AB}}$ quantifiers.

\subsection{Dirac quantum fields in SST}
\label{sec3}
In asymptotically flat spacetime, the Dirac equation represents a spinor field $\chi$ with a wave function, as shown in \cite{brill1957interaction, he2015property}:
\begin{equation}
	[i\gamma^ae^\mu_a(\partial_\mu +
	\Gamma_\mu)]\chi = 0,
	\label{eq6}
\end{equation}
The symbol $\gamma^a$ denotes the Dirac gamma matrix, $e^\mu_a$ represents the inverse of the tetrad $e_\mu^a$, and $\Gamma_\mu$ is equal to $\frac{1}{8}\left[\gamma^a,\,\gamma^b \right]$. The expression $e^\nu_ae_{b\nu,\mu}$ denotes the spin connection coefficient. In our further discussion, we will adopt natural units by setting the values of $G$, $c$, $\hbar$, and $k_B$ to be equal to 1. In addition, we will be examining the asymptotically flat SST, which may be characterized by the following metric element:
\begin{equation}
	ds^2=-(1-\frac{2M_{H}}{r} ) dt^2+(1-\frac{2M_{H}}{r})^{-1} dr^2+r^2(d\theta^2+\sin^2\theta d\alpha^2),
	\label{eq7}
\end{equation}
with $M_{H}$ denotes the BH mass, the Dirac equation in SST may be rewritten by combining Eq. (\ref{eq6}) with Eq. (\ref{eq7}).
\begin{eqnarray}\label{eq8}
	-\Xi\frac{\partial\chi}{\partial t}&+&\gamma_{1}\sqrt{1-\frac{2M_{H}}{r}}\left[\frac{\partial}{\partial r}+\frac{1}{r}+\frac{M_{H}}{2r(r-2M_{H})} \right]\chi\nonumber\\&+& \frac{\gamma_{2}}{r} (\frac{\partial}{\partial\theta}+\frac{\cot\theta}{2})\chi+\frac{\gamma_{3}}{r\sin\theta}\frac{\partial\chi}{\partial\alpha}=0.
\end{eqnarray}
with $\Xi=\frac{\gamma_{0}}{\sqrt{1-\frac{2M_{H}}{r}}}$. To solve the Dirac equation (\ref{eq8}) near the EH, it is essential to extend the spinor field $\chi$ by utilizing positive and negative frequency solutions for the outside and inside of the EH, respectively. You can find detailed descriptions of these solutions in \cite{xu2014hawking, xu2014probing, jing2004late, jing2004dirac}.
\begin{equation}
	\chi^{I+}_{\varsigma_i}(r>r_{+})=\mathfrak{P}e^{-i\omega_i\mathfrak{u}}, \qquad \chi^{II+}_{\varsigma_i}(r<r_{+})=\mathfrak{P}e^{i\omega_i\mathfrak{u}}, \quad \forall i,
	\label{eq9}
\end{equation}
where $\mathfrak{P}$ represents a 4-component Dirac spinor \cite{wang2010projective,deng2011hawking}, $\omega_i$ indicates the monochromatic frequency of the Dirac field, and $\mathfrak{u}=t-r-2M_{H}\ln\left[\frac{r-2M_{H}}{2M_{H}} \right]$ acts as the tortoise coordinate \cite{he2016measurement}. Since $\chi^{I+}_{\varsigma_i}$ and $\chi^{II+}_{\varsigma_i}$ are analytic outside and inside the EH respectively, they form a complete orthogonal basis. Here, $I$ is the exterior region and $II$ is the interior region. Therefore, the Dirac field $\chi_{out}$ can be quantized in SST.
\begin{equation}
	\chi_{out}=\sum_{i,\vartheta}\int dk\left(a^{\vartheta}_{\varsigma_i}\chi^{\vartheta+}_{\varsigma_i}+b^{\vartheta+}_{\varsigma_i}\chi^{\vartheta-}_{\varsigma_i} \right) ,
	\label{eq10}
\end{equation}
where $a^{\vartheta}_{\varsigma_i}$ and $b^{\vartheta+}_{\varsigma_i}$ represent the fermionic annihilation operator and anti-fermionic creation operator, respectively, with $\vartheta=(I, II)$. On the other hand, the viewpoint of an accelerating observer in an asymptotically flat space-time may be correctly characterized using the Kruskal coordinates \cite{wang2010entanglement}. Similarly, for SST, the generalized light-like Kruskal coordinates, denoted by $\mathfrak{U}$ and $\mathfrak{V}$, are introduced as
\begin{widetext}
	\begin{minipage}{\linewidth}
\begin{equation}
	\mathfrak{u}=-4M_{H}\ln\left[ \frac{-\mathfrak{U}}{4M_{H}}\right], \qquad \mathfrak{v}=4M_{H}\ln\left[ \frac{\mathfrak{V}}{4M_{H}}\right], \quad \text{if}\quad r>r_{+}, 
	\label{eq11}
\end{equation}

\begin{equation}
	\mathfrak{u}=-4M_{H}\ln\left[ \frac{\mathfrak{U}}{4M_{H}}\right], \qquad \mathfrak{v}=4M_{H}\ln\left[ \frac{\mathfrak{V}}{4M_{H}}\right], \quad \text{if}\quad r<r_{+}, 
	\label{eq12}
\end{equation}
	\end{minipage}
\end{widetext}
Based on \cite{damour1976black}, it is possible to perform an analytic continuation for $\chi^{I+}_{\varsigma_i}$ and $\chi^{II+}_{\varsigma_i}$. Furthermore, it is possible to find the complete bases for positive energy modes that are analytic for all real values of $\mathfrak{U}$ and $\mathfrak{V}$. Specifically,
\begin{equation}
	\Upsilon^{I+}_{\varsigma_i}=e^{2\pi M_{H}\omega_i}\chi^{I+}_{\varsigma_i}+e^{-2\pi M_{H}\omega_i}\chi^{II-}_{\varsigma_i},
	\label{eq13}
\end{equation}
\begin{equation}
	\Upsilon^{II+}_{\varsigma_i}=e^{-2\pi M_{H}\omega_i}\chi^{I-}_{-\varsigma_i}+e^{2\pi M_{H}\omega_i}\chi^{II+}_{\varsigma_i},
	\label{eq14}
\end{equation}
The Dirac field in Kruskal spacetime can then be decomposed using the aforementioned complete basis.
\begin{equation}
	\chi_{out}=\sum_{i,\vartheta}\int dk\left[2\cosh(4\pi M_{H}\omega_i) \right]^{-1/2}\left(c^{\vartheta}_{\varsigma_i}\Upsilon^{\vartheta+}_{\varsigma_i}+d^{\vartheta+}_{\varsigma_i}\Upsilon^{\vartheta-}_{\varsigma_i}\right),
	\label{eq15}
\end{equation}
where $c^{\vartheta}_{\varsigma_i}$ and $d^{\vartheta+}_{\varsigma_i}$ represent the annihilation and creation operators that act on the Kruskal vacuum. It is evident that the two quantization procedures shown in Eqs. (\ref{eq10}) and (\ref{eq15}) are not the same. The relationship between the creation and annihilation operators in the Schwarzschild and Kruskal coordinates is reflected in the Bogoliubov transformations \cite{barnett2002methods}. Since the Bogoliubov transformations are diagonal, we can express each annihilation operator \( c^{I}_{\varsigma_i} \) as a combination of Schwarzschild particle operators with a single Schwarzschild frequency \( \omega_i \) \cite{bruschi2010unruh, martin2010unveiling}. More specifically,

\begin{equation}
	c^{I}_{\varsigma_i}=\left(e^{-8\pi M_{H}\omega_i}+1 \right)^{-1/2}a^{I}_{\varsigma_i} - \left(e^{8\pi M_{H}\omega_i}+1 \right)^{-1/2}b^{II+}_{\varsigma_i}.
	\label{eq16}
\end{equation}
The vacuum and excited states of the Kruskal particle for mode $\Lambda$ in the Schwarzschild Fock space bases can be determined through a series of calculations. The representations of these states are as follows: $\left| 1\right\rangle_{\Lambda}=\otimes_i\left| 1_{\omega_i}\right\rangle_{\Lambda}$ and $\left| 0\right\rangle_{\Lambda}=\otimes_i\left| 0_{\omega_i}\right\rangle_{\Lambda}$.
\begin{widetext}
	\begin{minipage}{\linewidth}
\begin{equation}
	\begin{array}{cc}
		\begin{aligned}	
			\left| 0\right\rangle_{\Lambda}& =\left(e^{-\omega_i/T_H}+1 \right)^{-1/2} \exp\left[e^{-\omega_i/2T_H}a^{I+}_{\varsigma_i}b^{II+}_{-\varsigma_i} \right] \left| 0_{\varsigma_i}\right\rangle_I^+\left| 0_{-\varsigma_i}\right\rangle_{II}^-\\
			&=\left(e^{-\omega_i/T_H}+1 \right)^{-1/2}\left| 0_{\varsigma_i}\right\rangle_I^+\left| 0_{-\varsigma_i}\right\rangle_{II}^{-}+\left(e^{\omega_i/T_H}+1 \right)^{-1/2}\left| 1_{\varsigma_i}\right\rangle_I^+\left| 1_{-\varsigma_i}\right\rangle_{II}^{-}
		\end{aligned}
	\end{array},
	\label{eq17}
\end{equation}
	\end{minipage}
\end{widetext}
and 
\begin{equation}
	\left|1\right\rangle_{\Lambda}=c^{I+}_{\varsigma_i}\left| 0\right\rangle_{\Lambda}=\left| 1_{\varsigma_i}\right\rangle_I^+\left| 0_{-\varsigma_i}\right\rangle_{II}^{-},
	\label{eq18}
\end{equation}
where $T_H=\frac{1}{8\pi M_{H}}$ represents the Hawking temperature, we simplify $\left\lbrace\left| p_{\varsigma_i}\right\rangle^{+} \right\rbrace $ by $\left\lbrace\left| p\right\rangle_{I} \right\rbrace $ and $\left\lbrace\left| p_{-\varsigma_i}\right\rangle^{-} \right\rbrace $ by $\left\lbrace\left| p\right\rangle_{II} \right\rbrace $. These represent the orthonormal bases for the outside and inside of the EH, respectively. We note that in a fully quantum framework, the processes of BH formation and evaporation involve system entanglement. The definitions of the Kruskal particle's vacuum state (\ref{eq17}) and excited state (\ref{eq18}) clarify this idea. Furthermore, in the context of quantum information theory, Hawking decoherence can also be understood as a quantum noisy channel \cite{mancini2014preserving, bradler2009private}. For example, in Eqs. (\ref{eq17}) and (\ref{eq18}), the Hawking decoherence functions as an amplification quantum channel and is associated with the observable $\left(e^{\omega_i/T_H}+1 \right)^{-1/2}$\cite{yao2014quantum,aspachs2010optimal}.

\begin{widetext}
	\begin{minipage}{\linewidth}
\begin{figure}[H]
	\centering
	\includegraphics[width=0.5\linewidth]{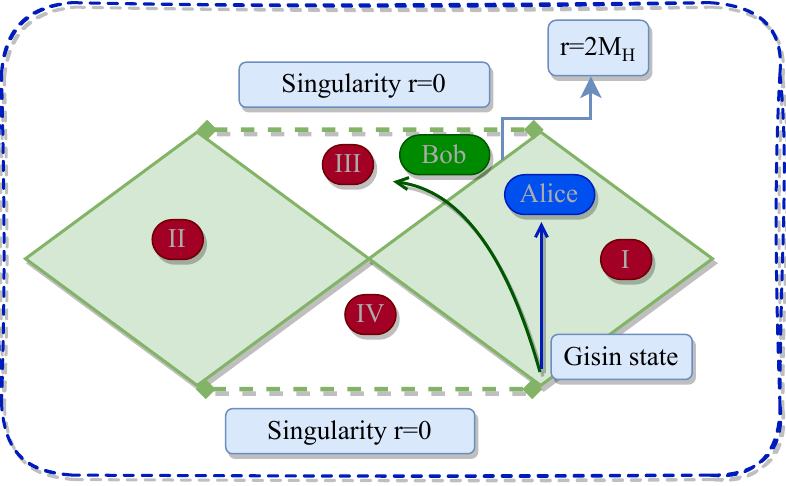}
	\caption{A Penrose diagram is useful for illustrating the trajectories of \textit{Bob} and \textit{Alice} in a SBH. It shows how they share an initial \textit{Gisin} state in the asymptotically flat region of SST. In this diagram, \textit{Bob} is shown as free-falling towards the SBH, while \textit{Alice} remains static. The interior of the BH is represented by region $III$, and the exterior is divided into regions $I$ and $II$. Region $IV$ represents the location of the white hole.}
	\label{fig:b}
\end{figure}
	\end{minipage}
\end{widetext}
We will investigate the quantum correlations of Dirac fields near a SBH, both in physically accessible and inaccessible regions, by examining the behavior of Dirac particles. From an inertial standpoint, the particle exists in a distinct superposition of Kruskal monochromatic modes. In SST, the observer detects the field in states defined by a single mode. For our analysis, let's consider two observers, \textit{Bob} and \textit{Alice}. Initially, they are both in an asymptotically flat region of SST, sharing an entangled \textit{Gisin} state. \textit{Bob}'s detector is sensitive to mode $\left|p\right\rangle_B$, while \textit{Alice}'s detector is intended to detect mode $\left|p\right\rangle_A$. After their first encounter, \textit{Alice} remains in the flat region that approaches infinity, whereas \textit{Bob} freely falls towards a SBH with constant acceleration, as shown in Figure (\ref{fig:b}), eventually arriving near its EH. To describe \textit{Bob}'s point of view, we must expand the initial \textit{Gisin} state corresponding to mode $\Lambda$, taking into account both the interior and exterior regions of the EH. The Kruskal states $\left|0 \right\rangle_{\Lambda} $ and $\left|1 \right\rangle_{\Lambda} $ correspond to two-mode states within the SST, as described by Eqs. \eqref{eq17} and \eqref{eq18}. By rewriting the initial \textit{Gisin} state, we obtain the evolved state.
\begin{widetext}
	\begin{minipage}{\linewidth}
\begin{eqnarray}
	\varrho_{AB_{I}B_{II}} &=& \frac{1-\lambda}{2}\left( \varpi^{2} |000\rangle\langle000|+\varpi\varepsilon(|000\rangle\langle011|+|011\rangle\langle000|)+\varepsilon^{2}|011\rangle\langle011|+|110\rangle\langle110|\right) \nonumber\\&+&\lambda\cos^{2}\psi\left(\varpi^{2} |100\rangle\langle100|+\varpi\varepsilon(|100\rangle\langle111|+|111\rangle\langle100|+\varepsilon^{2}|111\rangle\langle111|) \right)+ \lambda\sin^{2}\psi|010\rangle\langle010|\nonumber\\&+&\lambda\sin\psi\cos\psi\left(\varpi(|010\rangle\langle100|+|100\rangle\langle010|)+\varepsilon(|010\rangle\langle111|+|111\rangle\langle010|) \right). 
	\label{eq20}
\end{eqnarray}
	\end{minipage}
\end{widetext}
where we define \(\varpi = \frac{1}{\left(e^{-\omega/T_H} + 1\right)^{1/2}}
\), \(\varepsilon =\frac{1}{\left(e^{\omega/T_H} + 1\right)^{1/2}}\), and \(|opq\rangle = |o\rangle_A |p\rangle_{B_I} |q\rangle_{B_{II}}\). By tracing over one degree of freedom in the density operator $\varrho_{AB_{I}B_{II}}$, three subsystems can be obtained: a physically accessible subsystem \(\varrho_{AB_{I}}\), a physically inaccessible subsystem \(\varrho_{AB_{II}}\), and a spacetime subsystem \(\varrho_{B_{I}B_{II}}\). 
\section{Results and Discussion \label{sec4}}
Following the above setups, we now study the quantum correlations of Dirac fields that are present in every subsystem. Although the established global quantum state consists of fermionic modes both inside and outside the EH, it is a pure state. However, due to the causal disconnection between the interior and exterior regions, an observer or detector outside the SBH cannot access all the information. By using the partial trace operation, one can distinguish between physically accessible and inaccessible scenarios. Despite the inaccessibility of certain scenarios to experimental observation or detection, the global quantum state remains unitary and pure. With the prevailing consensus among physicists, including Hawking, now in favor of the conservation of quantum information, we can theoretically investigate both accessible and inaccessible scenarios. The following subsections will explore these scenarios in detail.
\subsection{Physically accessible scenario}
As previously mentioned, the exterior and interior regions of SBH are completely disconnected. Therefore, the information can only be accessed within the subsystem $\varrho_{AB_{I}}$. Within this subsystem, the corresponding reduced density operator is as follows:
\begin{eqnarray}
	\varrho_{AB_{I}} &=& \frac{1-\lambda}{2}\varpi^{2} |00\rangle\langle00|+(\frac{1-\lambda}{2}\varepsilon^{2}+\lambda\sin^{2}\psi)|01\rangle\langle01| \nonumber\\&+&\lambda\cos^{2}\psi\varpi^{2}|10\rangle\langle10|+\lambda\sin\psi\cos\psi\varpi(|01\rangle\langle10|+|10\rangle\langle01|)\nonumber\\&+&(\frac{1-\lambda}{2}+\lambda\cos^{2}\psi\varepsilon^{2})|11\rangle\langle11|.
	\label{eq21}
\end{eqnarray}

\begin{widetext}
	\begin{minipage}{\linewidth}
\begin{figure}[H]
	\centering
	\subfigure[]{\label{fig1}\includegraphics[scale=0.6]{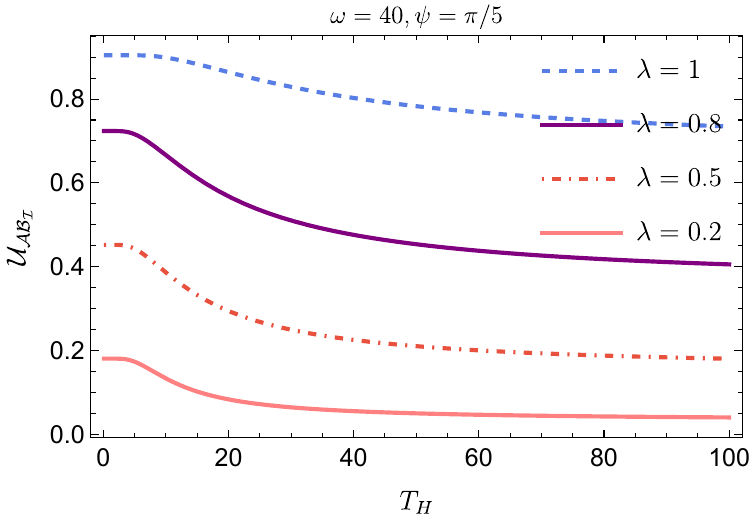}}
	\subfigure[]{\label{fig2}\includegraphics[scale=0.6]{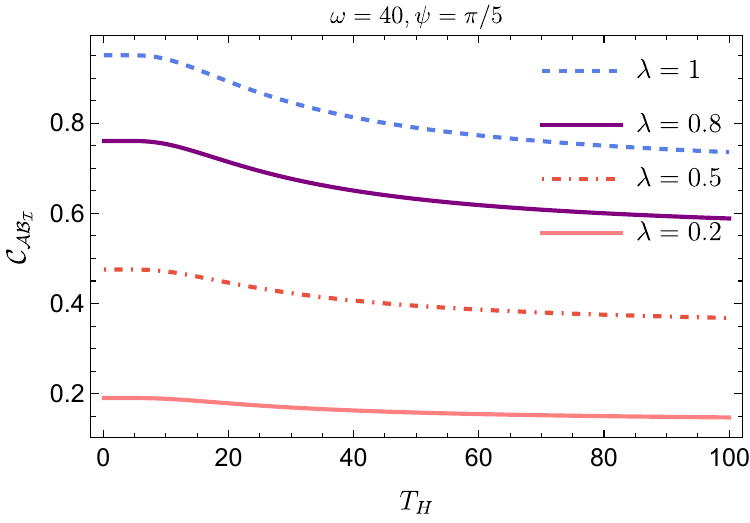}}
	\subfigure[]{\label{fig3}\includegraphics[scale=0.6]{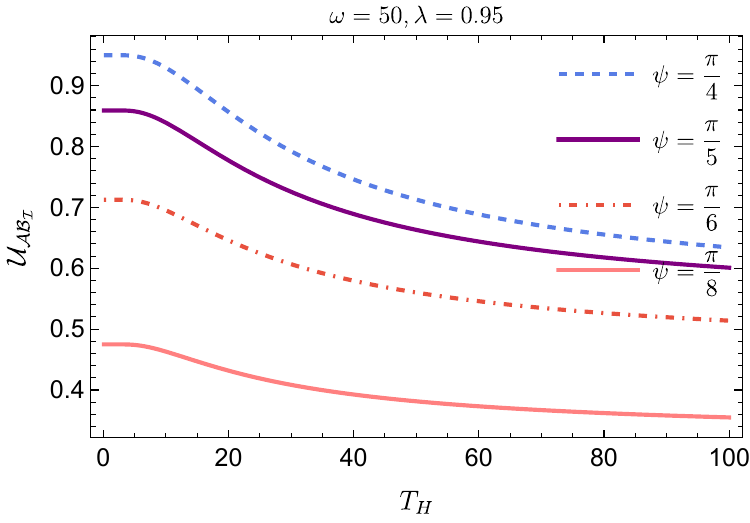}}
	\subfigure[]{\label{fig4}\includegraphics[scale=0.6]{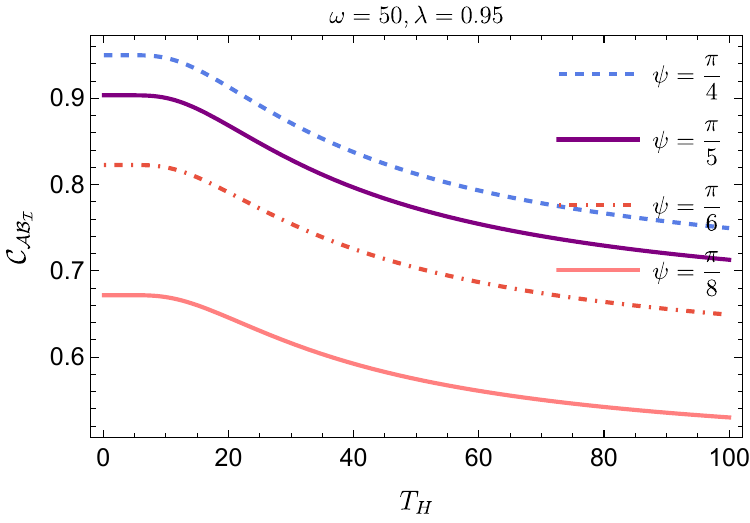}}
	\caption{Quantum correlations of Dirac fields $\mathcal{U_{AB_I}}$ \ref{fig1}--\ref{fig3} and $\mathcal{C_{AB_I}}$ \ref{fig2}--\ref{fig4} as a function of the Hawking temperature $T_H$ for different state parameters $\lambda$ and $\psi$ in the physically accessible system $\varrho_{AB_{I}}$.}
	\label{figure1}
\end{figure}
	\end{minipage}
\end{widetext}
To investigate the impact of various factors on $\mathcal{U_{AB_I}}$ and $\mathcal{C_{AB_I}}$, namely the Hawking temperature $T_H$, the \textit{Gisin} state parameters $\lambda$ and $\psi$, we analyze their dynamic behaviors in relation to $T_H$. Figure (\ref{figure1}) demonstrates how $T_H$ affects $\mathcal{U_{AB_I}}$ and $\mathcal{C_{AB_I}}$ while keeping other parameters, such as the mode frequency $\omega$, constant. Figures \ref{fig1}--\ref{fig4} consistently show a decrease and eventual disappearance of quantum correlations as $T_H$ approaches infinity. At $T_H=0$, quantum correlations reach their maximum and remain unaffected by the Hawking effect. However, as $T_H$ increases, decoherence occurs due to the interaction between the \textit{Gisin} state system and Hawking decoherence. The strength of the quantum correlations depends on the state parameters $\lambda$ and $\psi$, even at finite $T_H$. Reducing the mixing parameter $\lambda$ decreases both $\mathcal{U_{AB_I}}$ and $\mathcal{C_{AB_I}}$, regardless of the presence of the Hawking effect. Similarly, Figures \ref{fig3}--\ref{fig4}, illustrate that when $\psi = \frac{\pi}{4}$, both $\mathcal{U_{AB_I}}$ and $\mathcal{C_{AB_I}}$ reach their maximum values, with or without the Hawking effect. The values of $\mathcal{U_{AB_I}}$ and $\mathcal{C_{AB_I}}$ for Dirac particles close to a SBH are determined by the state parameters $\lambda$ and $\psi$. Increasing $\lambda$ enhances the quantum correlations, while increasing the Hawking temperature $T_H$ gradually suppresses them. When $T_H$ approaches infinity, corresponding to the complete evaporation of the BH, residual quantum correlations only exist for higher values of $\lambda$. As $\psi$ increases, the values of $\mathcal{U_{AB_I}}$ and $\mathcal{C_{AB_I}}$ also increase. However, further increasing $\psi$ from $\frac{\pi}{8}$ to $\frac{\pi}{4}$ results in an increase in these measures, ultimately leading to a separable classical state at $\psi=\frac{\pi}{8}$. In addition, as $T_H$ increases, quantum correlations in Figures \ref{fig3}--\ref{fig4} deteriorate rapidly compared to Figures \ref{fig1}--\ref{fig2}, shifting towards the right. The figures illustrate a linear increase in quantum correlations with increasing values of $\lambda$ and $\psi$. Moreover, further increases in $\lambda$ and $\psi$ lead to a gradual enhancement of quantum correlations, peaking at $\lambda = 0.9$. This finding aligns with \cite{elghaayda2024distribution}, suggesting that quantum correlations in this region serve as a resource for quantum information tasks in SST.

\begin{widetext}
	\begin{minipage}{\linewidth}
\begin{figure}[H]
	\centering
	\subfigure[]{\label{fig5}\includegraphics[scale=0.6]{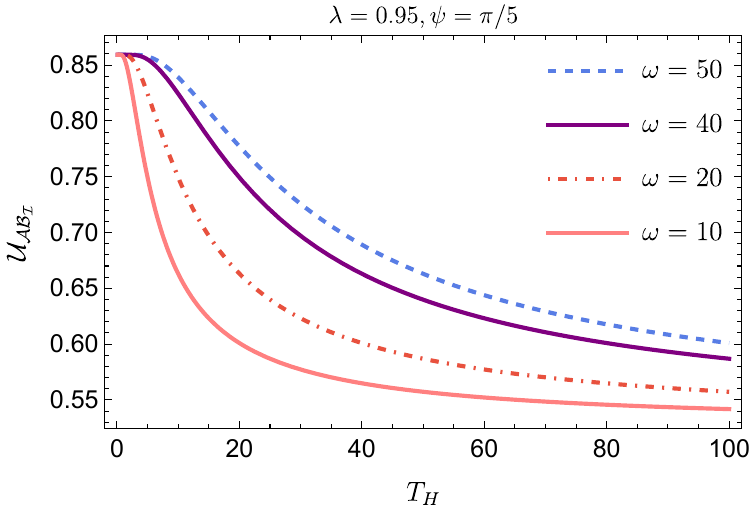}}
	\subfigure[]{\label{fig6}\includegraphics[scale=0.6]{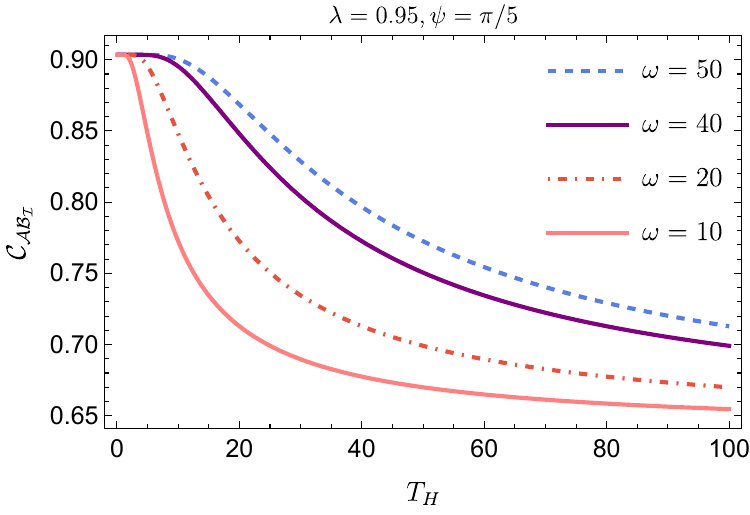}}
	\caption{Quantum correlations of Dirac fields $\mathcal{U_{AB_I}}$ \ref{fig5} and $\mathcal{C_{AB_I}}$ \ref{fig6} as a function of the Hawking temperature $T_H$ for different frequency modes $\omega$ in the physically accessible system $\varrho_{AB_{I}}$. }
	\label{figure2}
\end{figure}
	\end{minipage}
\end{widetext}
The effect of different frequencies $\omega$ of observed or detected Dirac particles on $\mathcal{U_{AB_I}}$ and $\mathcal{C_{AB_I}}$ is shown in Figure \ref{figure2}, while keeping all other parameters constant. As shown in Figure \ref{figure2}, the quantum correlations that can be accessed always decrease as $T_H$ increases. The presence of Hawking decoherence seems to speed up this reduction, potentially even leading to the sudden death of $\mathcal{U_{AB_I}}$ under certain conditions. In the limit $T_H \rightarrow \infty$, which corresponds to a BH with negligible mass, $\mathcal{U_{AB_I}}$ has a degree of zero, indicating a local state. For any finite Hawking temperature, the state becomes nonlocal, with lower Hawking temperatures resulting in greater nonlocal quantum correlations. As $T_H \rightarrow 0$, which corresponds to no BH evaporation, the degree of quantum correlations approaches 0.9, indicating a maximally correlated state. This demonstrates that only Dirac particles with high frequencies ($\omega \rightarrow \infty$) exhibit high correlations. Specifically, $\mathcal{C_{AB_I}}$ in a physically accessible region does not approach zero, regardless of the evaporation state of the BH, with a minimum value of 0.65 for $\omega = 10$. Additionally, the Pauli exclusion principle in Fermi-Dirac statistics prevents the quantum states of Fermi particles from being infinitely excited by the Hawking effect. This explains why the accessible correlations $\mathcal{C_{AB_I}}$ do not completely vanish, even when the Hawking temperature increases indefinitely. As $\omega$ decreases, the negative effects of Hawking decoherence become more noticeable. The maximum quantum correlations are observed for all frequencies at $T_H = 0$. However, a higher value of $\omega$ improves the resilience of the quantum correlations of the shared \textit{Gisin} state against Hawking decoherence. We observe that the value of $\mathcal{C_{AB_I}}$ is always nonzero. This indicates that $\mathcal{C_{AB_I}}$ is more robust than $\mathcal{U_{AB_I}}$ and can provide more information about the given system. In other words, $\mathcal{C_{AB_I}}$ can be considered a more effective resource for implementing RQI in practical applications.

\subsection{Physically inaccessible scenario}
We will now investigate how Hawking decoherence affects the properties of the shared \textit{Gisin} state in the physically inaccessible subsystem $\varrho_{AB_{II}}$. To comprehend the whereabouts of the lost quantum correlations, we assume that these vanished quantum correlations are redistributed to physically inaccessible regions. This assumption is based on the fact that Hawking decoherence splits Bob's mode into mode $\left| p\right\rangle_{B_{I}}$ and mode $\left| p\right\rangle_{B_{II}}$. Our focus will be on the remaining two subsystems, where we will explore the quantum correlations in the physically inaccessible region. The subsystem $\varrho_{AB_{II}}$ is given by

\begin{eqnarray}
	\varrho_{AB_{II}} &=&(\frac{1-\lambda}{2}\varpi^{2}+\lambda\sin^{2}\psi)|00\rangle\langle00|+ \frac{1-\lambda}{2}\varepsilon^{2} |01\rangle\langle01|\nonumber\\&+&(\frac{1-\lambda}{2}+\lambda\cos^{2}\psi\varpi^{2})|10\rangle\langle10|+ \lambda\cos^{2}\psi\varepsilon^{2}|11\rangle\langle11|\nonumber\\&+&\lambda\sin\psi\cos\psi\varepsilon(|00\rangle\langle11|+|11\rangle\langle00|)
	\label{eq22}
\end{eqnarray}

\begin{widetext}
	\begin{minipage}{\linewidth}
\begin{figure}[H]
	\centering
	\subfigure[]{\label{fig7}\includegraphics[scale=0.6]{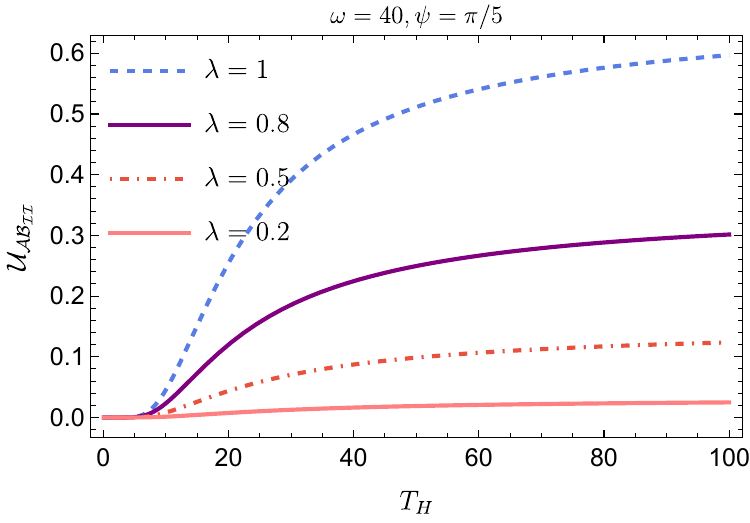}}
	\subfigure[]{\label{fig8}\includegraphics[scale=0.6]{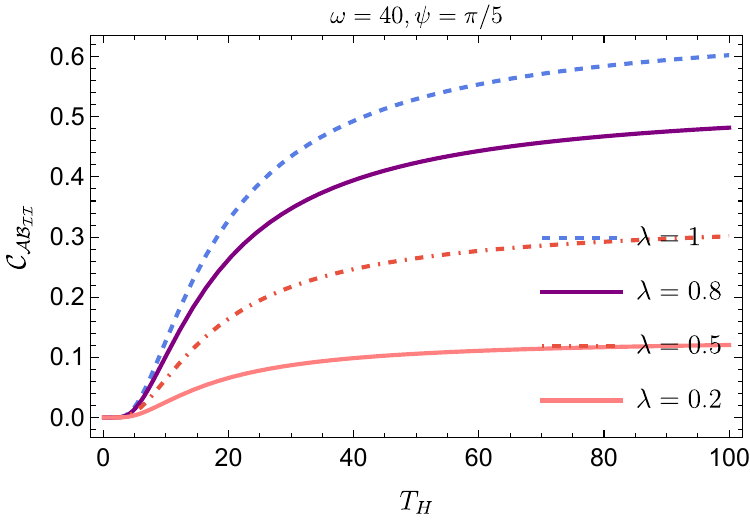}}
	\subfigure[]{\label{fig9}\includegraphics[scale=0.6]{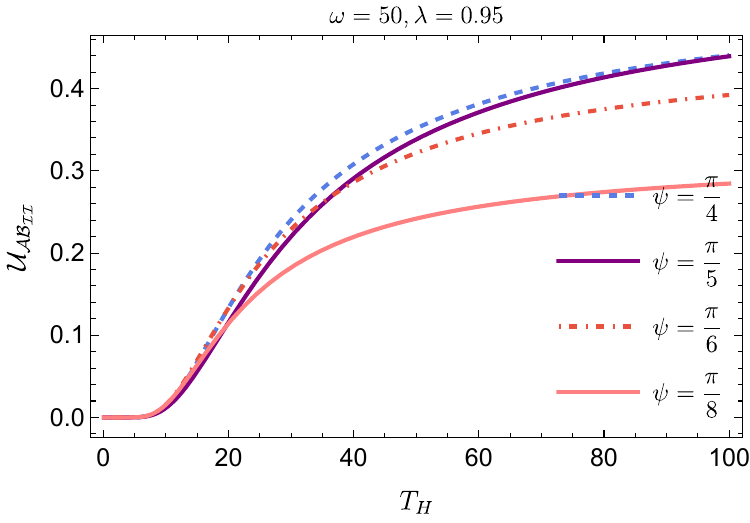}}
	\subfigure[]{\label{fig10}\includegraphics[scale=0.6]{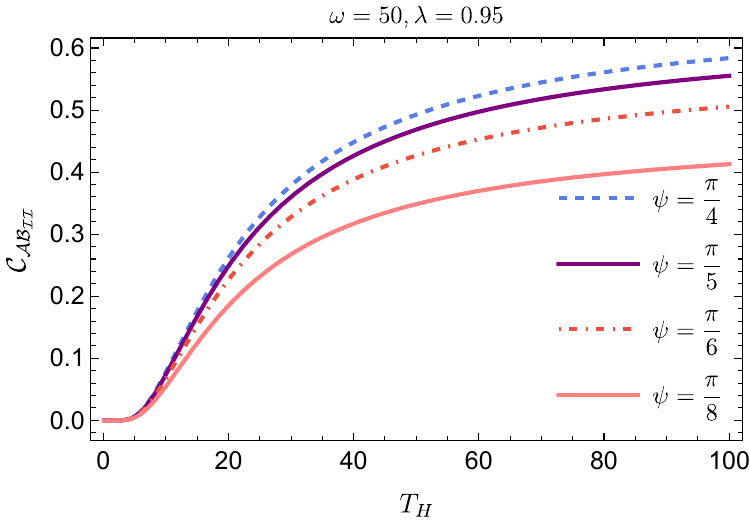}}
	\caption{Quantum correlations of Dirac fields $\mathcal{U_{AB_{II}}}$ \ref{fig7}--\ref{fig9} and $\mathcal{C_{AB_{II}}}$ \ref{fig8}--\ref{fig10} as a function of the Hawking temperature $T_H$ for different state parameters $\lambda$ and $\psi$ in the physically inaccessible system $\varrho_{AB_{II}}$.}
	\label{figure3}
\end{figure}
	\end{minipage}
\end{widetext}
In Figure \ref{figure3}, we plot $\mathcal{U_{AB_{II}}}$ and $\mathcal{C_{AB_{II}}}$ as functions of $T_H$ for different values of the state parameters $\lambda$ and $\psi$. The plot shows that in the physically inaccessible subsystem $\varrho_{AB_{II}}$, $\mathcal{U_{AB_{II}}}$ and $\mathcal{C_{AB_{II}}}$ are zero between modes $A$ and $B_{II}$ when Hawking decoherence is absent. When the Hawking temperature increases, the inaccessible quantum correlations monotonically increase to a nonzero value, in contrast to the $N$-partite entanglement which has a monotonic and non-monotonic relationship with the Hawking temperature \cite{wu2024genuinely}. It should be noted that the amount of quantum correlations depends on the initial \textit{Gisin} state parameters $\lambda$ and $\psi$. In other words, the initial state parameters control how quantum information travels between the interior and exterior regions of the SBH and how the quantum information is influenced by Hawking decoherence. Previous studies \cite{shi2018quantum,wu2024genuinely,elghaayda2024distribution} have shown that increasing the state parameters can mitigate the decoherence effect caused by Hawking radiation. However, within the inaccessible region, enhancing the state parameters $\lambda$ and $\psi$ can amplify the decay of quantum resources in the system under study. As shown in Fig. \ref{figure1}, it is evident that the Hawking decoherence environment in the accessible region leads to a higher loss of quantum correlations compared to the Hawking decoherence environment in the inaccessible region. 

\begin{widetext}
	\begin{minipage}{\linewidth}
\begin{figure}[H]
	\centering
	\subfigure[]{\label{fig11}\includegraphics[scale=0.6]{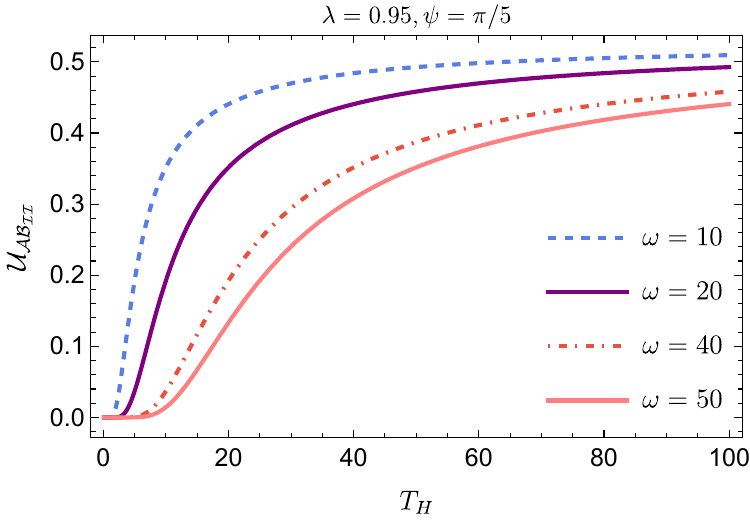}}
	\subfigure[]{\label{fig12}\includegraphics[scale=0.6]{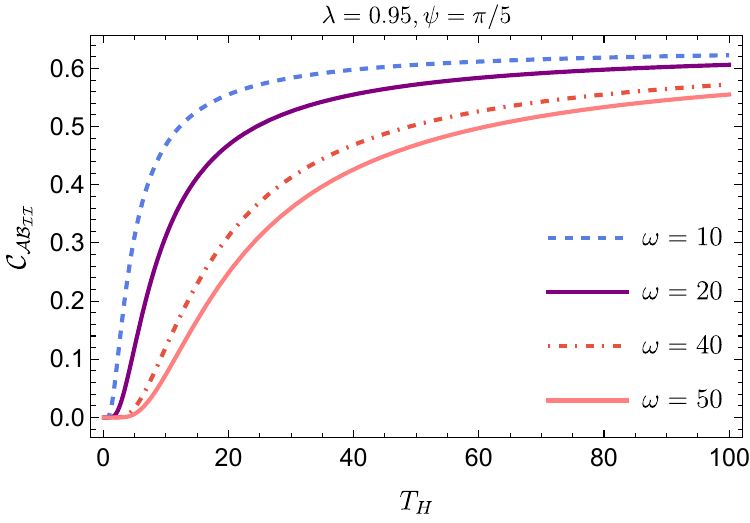}}
	\caption{Quantum correlations of Dirac fields
		$\mathcal{U_{B_{I}B_{II}}}$ \ref{fig11} and $\mathcal{C_{B_{I}B_{II}}}$ \ref{fig12} as a function of the Hawking temperature $T_H$ for different frequency modes $\omega$ in the spacetime region $\varrho_{B_{I}B_{II}}$. }
	\label{figure4}
\end{figure}
	\end{minipage}
\end{widetext}
We now analyze how the frequency mode of Dirac particles influences $\mathcal{U_{AB_I}}$ and $\mathcal{C_{AB_{II}}}$. To facilitate a direct comparison between these correlation measures, we present $\mathcal{U_{AB_I}}$ and $\mathcal{C_{AB_{II}}}$ as functions of the Hawking temperature in Figure (\ref{figure4}). In the absence of Hawking decoherence, there are no quantum correlations between modes $A$ and $B_{II}$, indicating that quantum correlations of Dirac fields remain finite at finite Hawking temperatures. At an infinite Hawking temperature ($T_H = 100$), the system's state $\varrho_{AB_{II}}$ shows partial correlation in the physically inaccessible region that can serve as a resource for specific RQI tasks. In contrast, for the physically accessible system $\varrho_{AB_{I}}$, quantum correlations approach zero under this limit. This behavior arises due to statistical correlations inherent in fermionic fields beyond entanglement, which persist and cannot be eliminated. This statistical correlation corresponds to the second quantized manifestation of statistical entanglement discussed in \cite{schliemann2001quantum}. Notably, these conclusions extend straightforwardly to the scenario of a Schwarzschild black hole. Therefore, employing the low-frequency mode ($\omega \approx 10$) preserves quantum correlations in this region, while the high-frequency mode is suitable for handling tasks related to RQI. Our observations of their distinct frequency dependencies in SST represent a novel contribution to understanding the Hawking decoherence of BHs.

\subsection{Spacetime scenario}
Considering the spacetime region, the reduced density operator in this region, $\varrho_{B_{I}B_{II}}$ is

\begin{eqnarray}
	\varrho_{B_{I}B_{II}} &=&(\frac{1-\lambda}{2}+\lambda\cos^{2}\psi)\varpi^{2}|00\rangle\langle00|+(\frac{1-\lambda}{2}+\lambda\cos^{2}\psi)\nonumber\\&\times&\varpi\varepsilon (|00\rangle\langle11|+|11\rangle\langle00|)+ (\frac{1-\lambda}{2}+\lambda\sin^{2}\psi)|10\rangle\langle10|\nonumber\\&+&(\frac{1-\lambda}{2}+\lambda\cos^{2}\psi)\varepsilon^{2}|11\rangle\langle11|
	\label{eq23}
\end{eqnarray}

\begin{widetext}
	\begin{minipage}{\linewidth}
\begin{figure}[H]
	\centering
	\subfigure[]{\label{fig13}\includegraphics[scale=0.6]{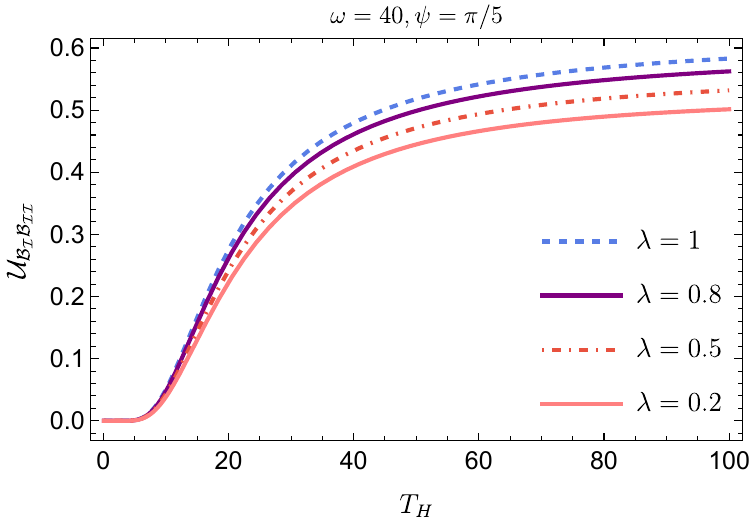}}
	\subfigure[]{\label{fig14}\includegraphics[scale=0.6]{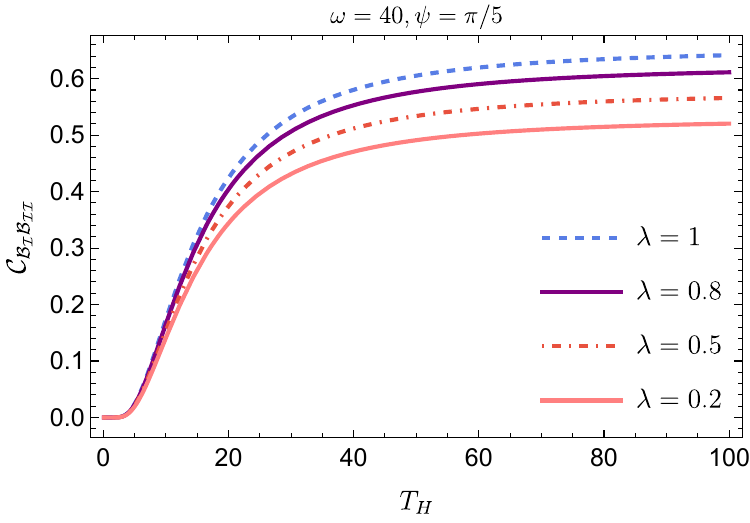}}
	\subfigure[]{\label{fig15}\includegraphics[scale=0.6]{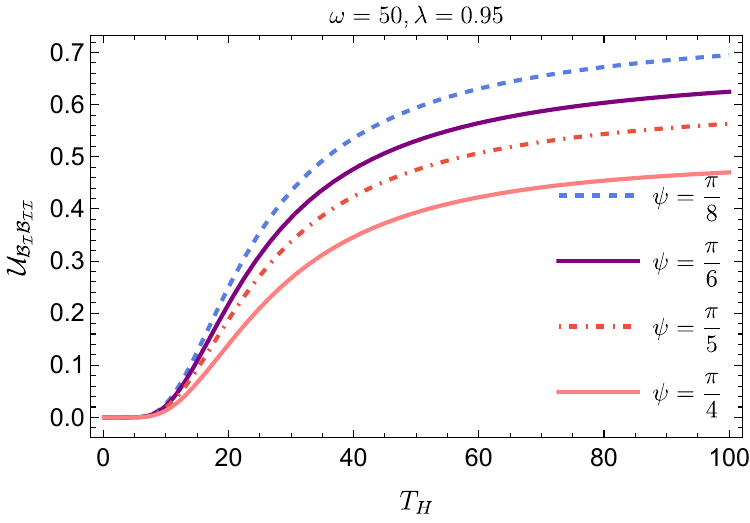}}
	\subfigure[]{\label{fig16}\includegraphics[scale=0.6]{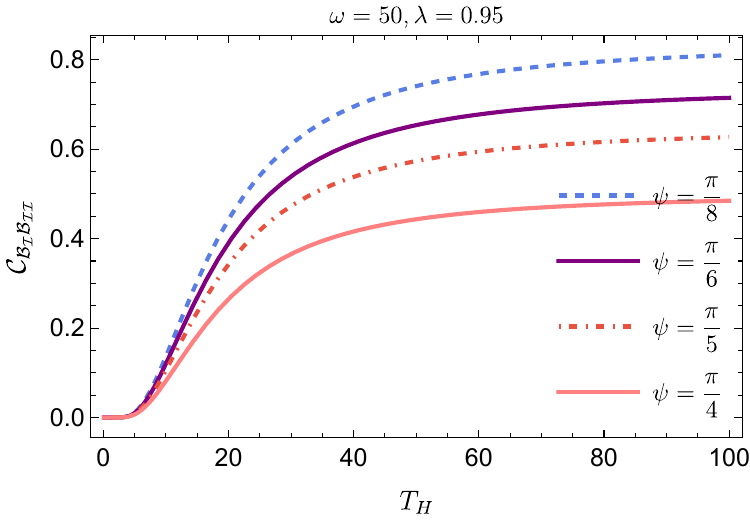}}
	\caption{Quantum correlations of Dirac fields 
		$\mathcal{U_{B_{I}B_{II}}}$ \ref{fig13}--\ref{fig15} and $\mathcal{C_{B_{I}B_{II}}}$ \ref{fig14}--\ref{fig16} as a function of the Hawking temperature $T_H$ for different state parameters $\lambda$ and $\psi$ in the space–time region $\varrho_{B_{I}B_{II}}$. }
	\label{figure5}
\end{figure}
	\end{minipage}
\end{widetext}
Here, we present a comprehensive examination of quantum correlations within the region of spacetime affected by Hawking decoherence. In Figure (\ref{figure5}), we illustrate the behaviors of $\mathcal{C_{B_{I}B_{II}}}$ and $\mathcal{U_{B_{I}B_{II}}}$ as a function of $T_H$, for various values of the state parameters $\lambda$ and $\psi$. It is clear that the quantum correlations reach their maximum level at higher Hawking temperatures. Within this region of spacetime, these quantum correlations are able to recover at zero Hawking temperature and gradually approach a constant value as the Hawking temperature approaches infinity. It is worth noting that the amount of quantum correlations decreases dramatically to zero at a temperature value $T_H\approx 10$, which depends on the \textit{Gisin} state parameters, and then increases monotonically to reach a stable value. This result suggests the possibility of Hawking-induced quantum correlations in this region of spacetime. Additionally, we observe that the quantum correlations increase as the parameter $\lambda$ increases, particularly for high Hawking temperatures ($T_H>40$). Therefore, the \textit{Gisin} state parameter $\lambda$ can be used to mitigate the detrimental effect of high Hawking temperature. Furthermore, from Figure (\ref{figure5}), we also observe that at the highest values of $T_H$ (i.e., $T_H>80$), $\mathcal{C_{B_{I}B_{II}}}$ is significantly larger than $\mathcal{U_{B_{I}B_{II}}}$, indicating that the current system is more coherent than correlated. In particular, based on Hawking's initial argument \cite{hawking1975particle,hawking1976breakdown}, larger black holes have a lower temperature, which results in weaker radiation emission compared to smaller black holes. Simply put, the mass of the black hole directly influences its radiation temperature. Our findings support Hawking's assertion by demonstrating that quantum correlations increase as the temperature of the black hole increases in the spacetime region. The maximum difference between $\mathcal{C_{B_{I}B_{II}}}$ and $\mathcal{U_{B_{I}B_{II}}}$ occurs as $T_H$ increases. In this case, $\mathcal{C_{B_{I}B_{II}}}$ sharply increases and then remains constant at an infinite value of $T_H$. On the other side, $\mathcal{U_{B_{I}B_{II}}}$ exhibits similar behavior. Furthermore, when the values of Hawking decoherence increase, $\mathcal{U_{B_{I}B_{II}}}$ experiences a significant increase followed by freezing.

\begin{widetext}
	\begin{minipage}{\linewidth}
\begin{figure}[H]
	\centering
	\subfigure[]{\label{fig17}\includegraphics[scale=0.6]{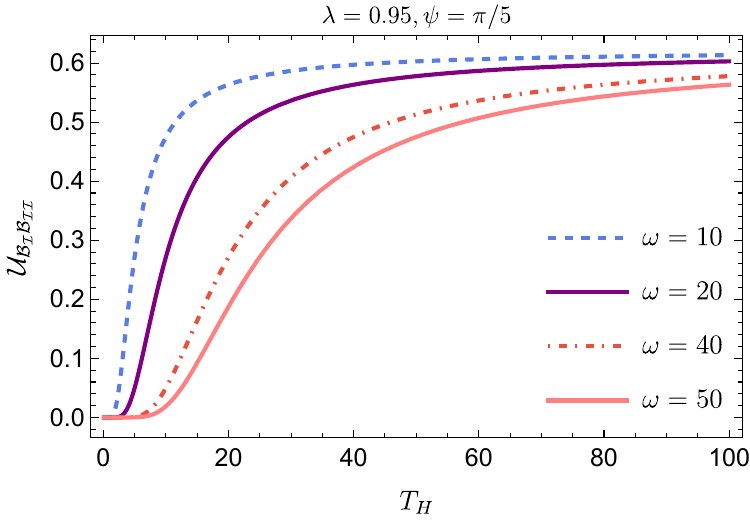}}
	\subfigure[]{\label{fig18}\includegraphics[scale=0.6]{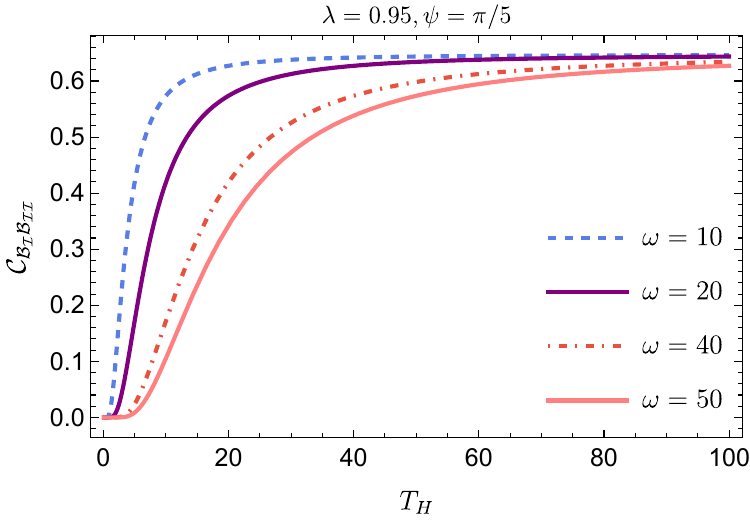}}
	\caption{Quantum correlations of Dirac fields
		$\mathcal{U_{B_{I}B_{II}}}$ \ref{fig17} and $\mathcal{C_{B_{I}B_{II}}}$ \ref{fig18} as a function of the Hawking temperature $T_H$ for different frequency modes $\omega$ in the space–time region $\varrho_{B_{I}B_{II}}$. }
	\label{figure6}
\end{figure}
	\end{minipage}
\end{widetext}
We now analyze the influence of the frequency mode $\omega$ of the fermionic particles on the change of quantum correlations between modes $B_{I}$ and $B_{II}$ in the presence of Hawking decoherence. In Figure (\ref{figure6}), $\mathcal{C_{B_{I}B_{II}}}$ and $\mathcal{U_{B_{I}B_{II}}}$ are displayed against $T_H$ for different values of the frequency mode $\omega$. From Figure (\ref{figure6}), one can observe that $\mathcal{C_{B_{I}B_{II}}}$ exhibits similar behavior to $\mathcal{U_{B_{I}B_{II}}}$, but they have different values. Initially, both measures show zero correlations in the spacetime region, and then increase to attain a saturation value for sufficiently large $T_H\approx 100$. However, as the frequency mode $\omega$ increases, the influence of the Hawking temperature $T_H$ on the quantum correlations attenuates. This suggests that by adjusting the frequency mode $\omega$, quantum correlations can be improved and preserved. Therefore, when preparing \textit{Gisin} states in the spacetime region with significant correlations, the advantageous impact of Hawking temperature $T_H$ must be taken into account. The decline of quantum correlations is more noticeable for finite Hawking temperatures. This indicates that low levels of external Hawking decoherence adversely affect the quantum correlations between modes $B_{I}$ and $B_{II}$ in the evolved Gisin state. Consequently, as the Hawking temperature increases and the mode frequency decreases, the system's state in the spacetime region becomes more correlated.
\subsection{Quantum correlations redistribution}
Based on the analysis presented above, it is clear that the process of Hawking decoherence in a SBH leads to the redistribution of quantum correlations. This process shows that Hawking decoherence decreases the presence of accessible quantum correlations while at the same time increasing the amount of inaccessible quantum correlations in either a monotonic or non-monotonic way. Therefore, in order to understand how quantum information is distributed within a SBH, both in its interior and exterior regions, we will examine the distribution of quantum correlations in Dirac fields across various regions. Our main goal will be to analyze how these correlations are distributed among all subsystems.
\begin{widetext}
	\begin{minipage}{\linewidth}
\begin{figure}[H]
	\centering
	\subfigure[]{\label{fig21}\includegraphics[scale=0.6]{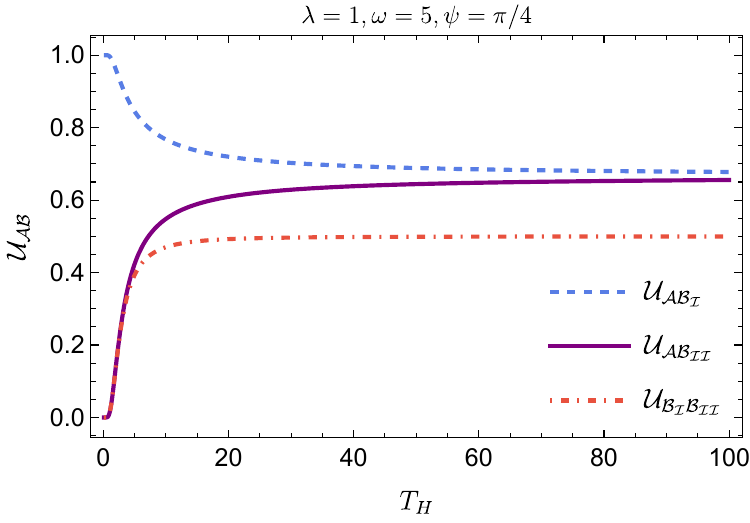}}
	\subfigure[]{\label{fig22}\includegraphics[scale=0.6]{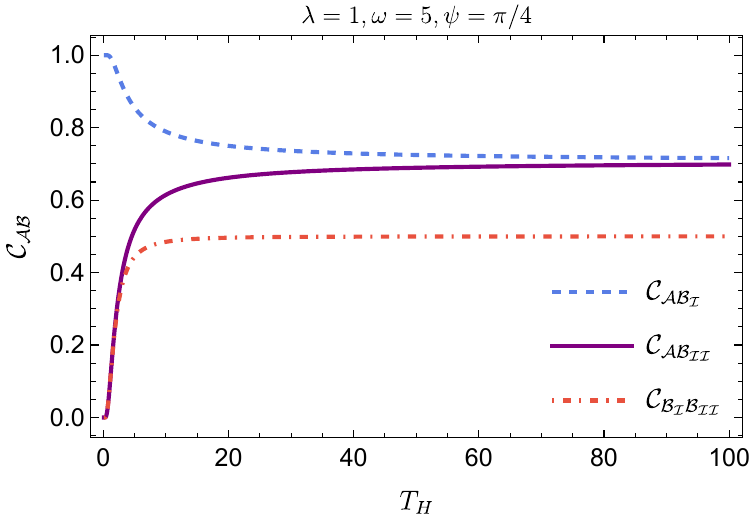}}
	\subfigure[]{\label{fig19}\includegraphics[scale=0.6]{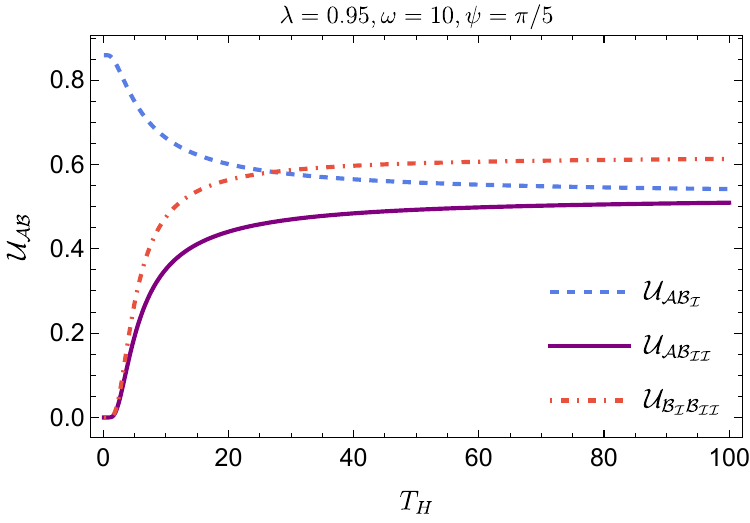}}
	\subfigure[]{\label{fig20}\includegraphics[scale=0.6]{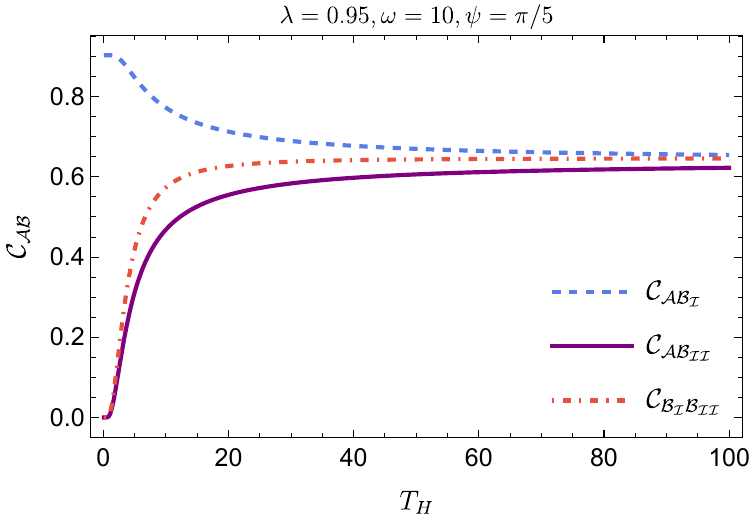}}
	\caption{Quantum correlations of Dirac fields
		$\mathcal{U_{AB}}$ \ref{fig21}--\ref{fig19} and $\mathcal{C_{AB}}$ \ref{fig22}--\ref{fig20} among all the subsystems $\varrho_{AB_{I}}$, $\varrho_{AB_{II}}$ and $\varrho_{B_{I}B_{II}}$ as functions of Hawking temperature $T_H$. }
	\label{figure7}
\end{figure}
	\end{minipage}
\end{widetext}
To thoroughly investigate the quantum correlations of Dirac fields in the SST, we plotted their distributions in Figure (\ref{figure7}). These distributions demonstrate the effect of Hawking decoherence on quantum correlations across all possible subsystems. We observe that the initially accessible quantum correlations decrease and then stabilize as the Hawking temperature $T_H$ increases. On the other hand, the initially inaccessible quantum correlations remain at zero and then gradually increase and stabilize as the Hawking decoherence intensifies. Our findings indicate that at low Hawking temperatures, mode $AB_{I}$ retains the majority of the quantum correlation, while the correlation is smaller in modes $AB_{II}$ and $B_{I}B_{II}$. As the Hawking decoherence increases, more accessible information leaks into the inaccessible regions, causing a greater degree of correlation between them. In essence, the original quantum information exchanged between \textit{Alice} and \textit{Bob} gets distributed across the modes $A$, $B_{I}$, and $B_{II}$. Furthermore, when the SBH completely evaporates, the quantum correlations of subsystems $\varrho_{AB_{I}}$ and $\varrho_{B_{I}B_{II}}$ will be approximately equal for a maximally entangled \textit{Gisin} state ($\lambda=1$ and $\psi=\pi/4$). Additionally, it is worth noting that physically accessible quantum correlations can be redistributed to a region that is physically inaccessible. Moreover, we have observed that the redistribution of quantum correlations does not occur in the same way when considering a mixed \textit{Gisin} state. This suggests that the maximally initially entangled \textit{Gisin} state is more effective in resisting Hawking decoherence and is better suited for processing RQI. These findings offer valuable guidance in choosing suitable quantum states and resources for RQI tasks. For example, the maximally initially entangled \textit{Gisin} state can be regarded as the quantum state, while quantum consonance can be chosen as the quantum resource.
\section{Concluding remarks \label{sec5}}
In conclusion, we have examined the distribution of physically accessible and inaccessible quantum correlations of Dirac fields within the SST framework. We have introduced two measures of quantum correlation: uncertainty-induced non-locality and quantum consonance. Through our analysis, we have demonstrated how quantum correlations of Dirac fields are distributed among different subsystems. Our findings indicate that an increase in Hawking radiation inevitably leads to a reduction in quantum correlations. This reduction can be explained as the leakage of information from the initial \textit{Gisin} state into SST, stimulated by Hawking decoherence. Specifically, the magnitudes of quantum correlation that are physically accessible are determined by the initial \textit{Gisin} state parameters and consistently diminish with intensifying Hawking radiation. In the physically inaccessible and spacetime regions, quantum correlations of Dirac fields are absent at zero Hawking temperature, but reemerge as Hawking decoherence increases, eventually approaching specific constants at infinite Hawking temperature. As a result, the growing quantum correlations of Dirac fields can be understood as the transfer of quantum information from the physically accessible to the inaccessible region. To provide a more intuitive understanding of the flow of information, we also examine their distributions among all subsystems. Our results suggest that in the absence of Hawking decoherence, mode $AB_{I}$ contains the highest amount of information, while modes $AB_{II}$ and $B_{I}B_{II}$ contain less information. As the intensity of Hawking decoherence increases, some of the accessible information leaks into regions that are inaccessible. Moreover, the quantum information in $\varrho_{AB_{I}}$ and $\varrho_{AB_{II}}$ is almost identical when the Hawking temperature approaches infinity. Finally, we anticipate that the results of our investigation will advance our understanding of the quantum information paradox in curved spacetime, particularly with regard to quantum correlations of Dirac fields.

\bibliography{references}
\bibliographystyle{unsrt}

\end{document}